\def\GRL {{\it Geophys.\ Res.\ Lett.\/}}
   
\def\JGR {{\it J. Geophys.\ Res.\/}}
\def\PF  {{\it Phys. Fluids.}}
\def\PRL {{\it Phys. Rev. Lett.}}

\def\PRep  {{\it Phys. Rep.}}
\def\PR  {{\it Phys. Rev.}}
\def\PSr {{\it Phys. Scr.}}
\def\NPG {{\it Nonlinear Processes in Geophys.}}

\def\erf {\mathop{\rm erf}\nolimits}
\def\sech {\mathop{\rm sech}\nolimits}

\documentstyle[aps,prl,epsf,twocolumn]{revtex}

\draft

\begin{document}

\wideabs{

\title{Trapped and Passing Electrons in BGK Solitary Waves}
\author{Li-Jen Chen and George K. Parks*}
\address{Physics Department, University of Washington, Seattle, WA 98195}
\address{*Also at Space Science Laboratory, University of California, Berkeley}

\maketitle

\begin{abstract}
This paper reexamines the physical roles of trapped and passing electrons
in Bernstein-Greene-Kruskal (BGK) electron 
solitary waves, also called the BGK electron holes (EH).
It is shown that the shielding of the positive core is
achieved by electrons trapped and oscillating inside
the potential energy trough, instead of the thermal screening
from the ambient plasma as previously thought.
The size of a BGK EH is therefore not
restricted to be greater than the Debye length $\lambda_D$.
\end{abstract}

}


In 1957, Bernstein, Greene and Kruskal \cite{BGK57} solved the one-dimensional,
time-independent Vlasov-Poisson equations and obtained the general solutions
for electrostatic nonlinear traveling waves, including solitary potential pulses.
Their derivation emphasized the special role played by the particles trapped
in the potential energy troughs. They demonstrated that one
could construct waves of arbitrary shapes by assigning the distribution of
trapped particles suitable for the desired wave form.

In 1967, Roberts and Berk \cite{Roberts67} provided a quasi-particle picture
for the electron phase space holes (EH) based on the results of a numerical
experiment on two-stream instability.
They followed the evolution of the electron phase space
boundary between $f=0$ and $f=1$
using the time-dependent Vlasov-Poisson equations with the
``water-bag'' model in which the distribution function only takes
discrete values, 0 and 1. They interpreted the elliptical empty
($f=0$) region associated
with a positive charge observed in the late stage of the nonlinear development
as a BGK EH. In order to explain the coalescence of neighboring EHs,
a negative effective mass was assigned to each EH to compensate for the
Coulomb repulsion of two positive EHs.
Thus, they suggested a quasi-particle picture that BGK EHs have positive charge
and negative mass. This picture is in use even today to interpret results in
computer simulations of EH disruptions
\cite{Saeki98}, and to model the mutual interaction of
electrostatic solitary waves in space plasma \cite{Krasovsky99}.

It was not until 1979 that BGK EHs were experimentally realized by the Ris\mbox{\o}
laboratory experiments \cite{Saeki79,Lynov79}. By applying large amplitude
potential pulses in a plasma-loaded wave guide, solitary potential pulses were excited,
including EHs and Korteweg-de Vries (KdV) solitons \cite{Washimi66}.
Investigations of the mutual interaction of EHs showed that two EHs
close to each other would coalesce if they have almost equal velocity and
they would pass through each other if their relative velocity was large \cite{Saeki79}.
The coalescence was interpreted in terms of the positive EH picture derived earlier
\cite{Saeki79,Lynov79}.

The analytical work that followed the Ris\mbox{\o} experiments mainly focused on constructing
solutions and obtaining the corresponding width-amplitude relations to facilitate the
comparison between BGK EHs and KdV solitons \cite{Turikov84,Schamel86,Lynov85}.
In addressing the quasi-particle picture of EHs,
Schamel \cite{Schamel86} concluded that EHs were positively charged, screened by the
ambient electrons over many Debye lengths ($\lambda_D$), and
had negative mass \cite{Schamel86}.
This conclusion supports the positive EH picture previously obtained from the numerical
experiments \cite{Roberts67,Berk70} and that
the minimum size of EHs of several $\lambda_D$ is a consequence of thermal screening by
the ambient electrons.

Turikov \cite{Turikov84} followed the BGK approach and constructed the trapped
electron distribution for a Maxwellian ambient electron distribution and
various solitary potential profiles. He restricted his study to
BGK EHs with phase space density zero in a hole center and the
results showed that the potential width increases with increasing amplitude.
This behavior is different from that of the KdV solitons whose width decreases
with increasing amplitude.
He also numerically simulated the temporal evolution of the EHs for different Mach
numbers to study the EH stability and found that EHs are quasi-stable
for Mach numbers less than 2.
One of the main conclusions that Turikov made was that EHs are purely
kinetic nonlinear objects in which trapped electrons play an important
role, but exactly what physical role trapped electrons played was not
addressed.

Space-borne experiments now show that electrostatic solitary waves (ESW) are
ubiquitous in Earth's magnetospheric
boundaries, shock, geomagnetic tail and auroral ionosphere
\cite{Bostrom88,Matsumoto94,Franz98,Ergun98,Ergun99,Bale98}.
While detailed properties of these solitary waves are continuously being
studied, it has been shown that in a number of cases the ESWs have
features that are consistent with BGK electron \cite{Muschietti99}
and ion \cite{Malkki89} mode solitary waves.
A statistical study by FAST satellite observations \cite{Ergun98,Ergun99} in the auroral
ionosphere has revealed that solitary pulses with a positive potential typically
have a Gaussian
half width ranging from less than one $\lambda_D$ to several $\lambda_D$ with
a mean of $1.80 \lambda_D$ and a standard deviation of $1.13 \lambda_D$.
However, the statistical analysis strongly favored large amplitude pulses
\cite{Ergun98,Ergun99} and smaller EHs if they existed were not sampled.
This question of how small EHs can be
is an important issue associated with how a collisionless plasma supports
nonlinear waves and needs to be further investigated.

To examine the physical roles played by trapped and passing
electrons, we start first with simple physical arguments and then
perform analytical
calculations using the same formulation used by Turikov \cite{Turikov84},
except that we relax the restriction of empty-centered EHs and obtain a more general
width-amplitude relation. The number density profiles of trapped and
passing electrons are calculated to quantify the separate contributions from
trapped and passing electrons to the charge density.
Implications of our results to the positive EH picture and minimum
size of BGK EHs are made.

\begin{figure}
\epsfxsize=8.0cm
\epsffile{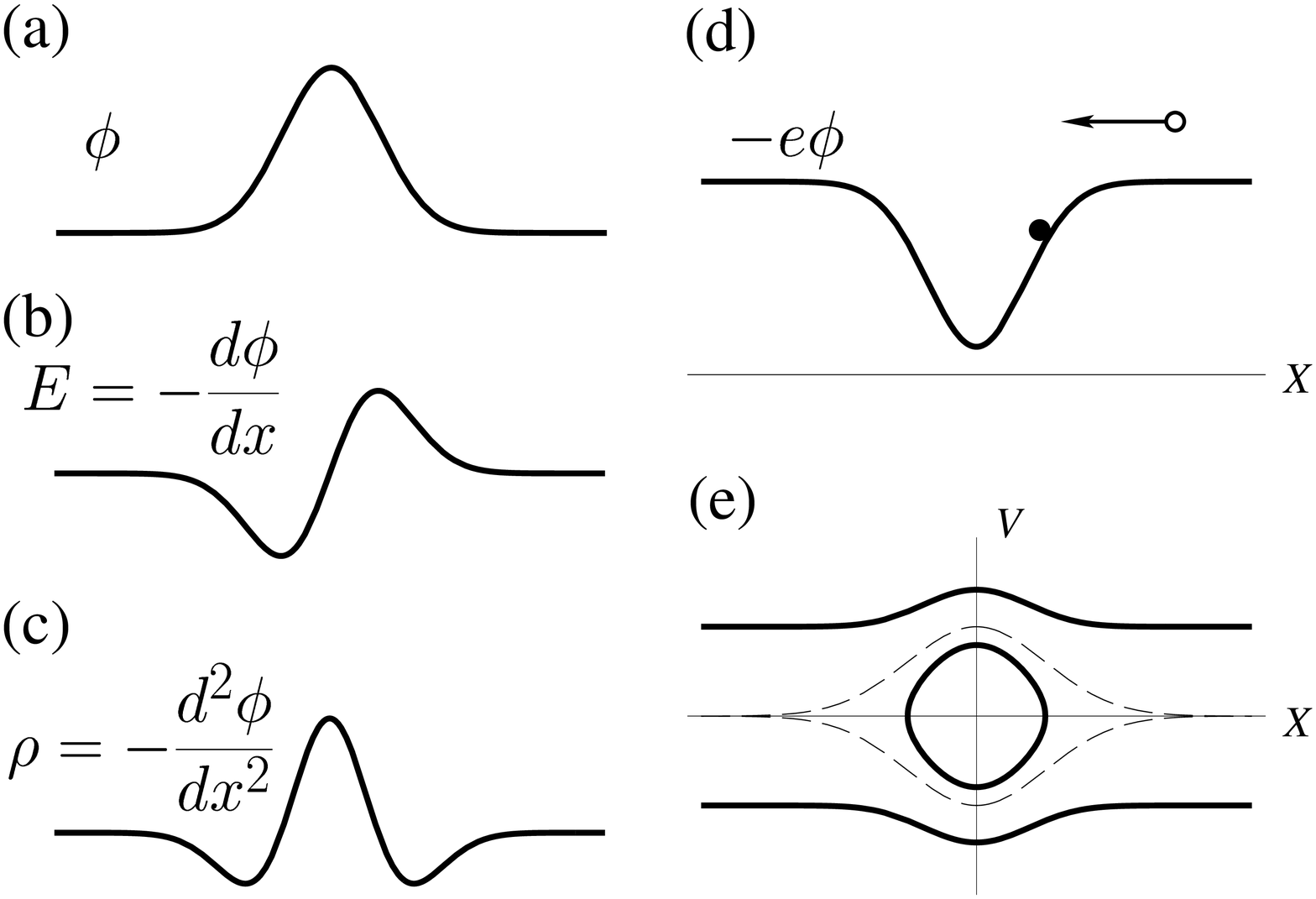}
\caption{Please see the text for explanations.}
\label{f1}
\end{figure}

We first discuss heuristically the behavior of
electrons in the vicinity of a potential pulse. Figures \ref{f1}(a)-\ref{f1}(c)
show the general form of a positive solitary potential pulse ($\phi(x)$),
the corresponding bipolar electric field ($E=-\partial \phi/\partial x$)
and the total charge density ($\rho=-\partial^2 \phi/\partial x^2$).
The charge
density is positive at the core, negative at the boundary, and zero outside
the solitary potential.
Figure \ref{f1}(d) shows the potential energy trough with an electron passing by
(open circle) and a trapped electron (solid circle) at its turning point.
Consider the phase space trajectories of electrons passing by the potential
and those that are trapped in the potential shown in Figure \ref{f1}(e).
The dashed line marked by electrons with zero total energy
is the boundary of the trapping region inside which electrons are trapped
and outside which electrons are untrapped.
The total energy, $w=\frac{m}{2}v^2-e\phi$, is a constant of motion.
A passing electron ($w>0$) moves with a constant velocity outside
the potential and the speed increases when it encounters the
potential pulse and then decreases back to its original value as it moves away.
A trapped electron ($w<0$) bounces back and forth between its two turning points
in the potential.
Since there is no source or sink for the particles, the density is inversely
proportional to the velocity.
We can thus deduce that the density of passing electrons is constant outside
and becomes
smaller as $\phi$ increases. No excess negative charge results from the
passing electrons. On the other hand, trapped electrons have density maxima
at their turning points, and so must be responsible for the excess negative
charge.
The charge density variation (Figure \ref{f1}(c)) needed to be self-consistent with
the solitary potential pulse is thus a net balance of the negative charge from
trapped electrons
and the positive charge due to the density decrease of passing electrons
since the ion density is assumed uniform.
From this simple picture, one can see that in BGK solitary waves it is
the trapped electrons traveling with the solitary potential that screen
out the positive core.
Our picture is different from the picture of a positive object in
a plasma whose screening is achieved by the thermal motion of the plasma
(Debye shielding).

The entire trapping region that consists of the total electron density
enhancement at the flanks and depletion at the core is a physical entity
produced by the self-consistent interaction between the plasma particles and
the solitary potential pulse. This defines the physical identity of one
BGK EH.
The total charge of the entire trapping region is zero, and therefore it
follows that two separated BGK EHs do not interact and the concept of
negative mass is not needed.

\begin{figure}
\epsfxsize=8.0cm
\epsffile{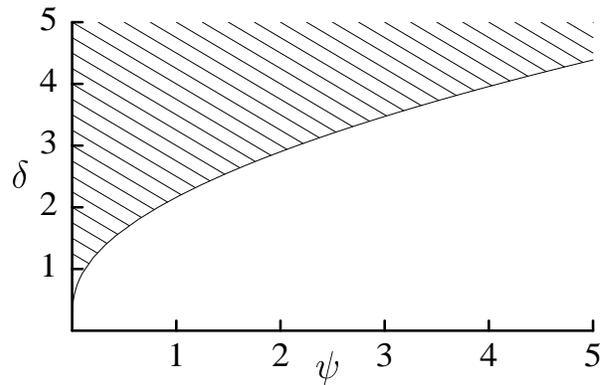}
\caption{the width-amplitude relation of BGK EHs that are not restricted to be empty-centered
for a Gaussian potential and Maxwellian ambient electron distribution}
\label{f2}
\end{figure}

We now use the approach formulated by BGK \cite{BGK57} to quantify
the above arguments
and further demonstrate that the results are independent of the strength of the
nonlinearity.
The time-independent, coupled Vlasov and Poisson equations with the assumption of a
uniform neutralizing ion background take the following form:
\begin{equation}
v\frac{\partial f(v, x)}{\partial x}+\frac{1}{2}\frac{\partial \phi}
{\partial x}\frac{\partial f(v, x)}{\partial v}=0, \label{Veq}
\end{equation}
\begin{equation}
\frac{\partial^2 \phi}{\partial x^2}=\int_{-\infty}^{\infty}f(v, x)dv-1,
\label{Peq}
\end{equation}
where f is the electron distribution function and the quantities have been
normalized with the units of the Debye length $\lambda_D$,
the ambient electron thermal energy $T_e$, and the electron thermal velocity
$v_t=\sqrt{2T_e/m}$.
The total energy $w=v^2-\phi$ under this convention.
Any $f=f(w)$ is a solution to Eq. \ref{Veq} as can be readily verified.
Recognizing this, Eq. \ref{Peq} can be
re-written in the following form,
\begin{equation}
\frac{\partial^2 \phi}{\partial x^2}=\int_{-\phi}^0 dw\frac{f_{tr}(w)}
{2\sqrt{w+\phi}}+\int_0^{\infty}dw \frac{f_p(w)}{2\sqrt{w+\phi}}-1, \label{phi2}
\end{equation}
where $f_{tr}(w)$ and $f_p(w)$ are the trapped and passing electron
phase space densities at energy $w$, respectively.
The first integral on the RHS of  Eq. \ref{phi2} is the trapped
electron density, and the second integral the passing electron density.
Prescribe the solitary potential as a Gaussian,
\begin{equation}
\phi(\psi, \delta, x)=\psi \exp{(-x^2/2\delta^2)},
\end{equation}
and the passing electron distribution a Maxwellian where
the density has been normalized to 1 outside the solitary potential,
\begin{equation}
f_p(w)=\frac{2}{\sqrt{\pi}}\exp{(-w)}.
\end{equation}
Following the BGK approach, we obtain the trapped electron distribution,
\begin{eqnarray}
f_{tr}(\psi, \delta, w) & = & \frac{4\sqrt{-w}}{\pi \delta^2}
\left[1-2\ln{(\frac{-4w}{\psi})}\right] \nonumber \\
 &  & +\frac{2\exp{(-w)}}{\sqrt{\pi}} \left[1-\erf(\sqrt{-w}) \right].
\label{ftr2}
\end{eqnarray}
The first term in $f_{tr}$ stems from $\partial^2\phi/\partial x^2$ term
in Eq. \ref{phi2} and has a single peak at $w=\frac{-\psi}{4e^{3/2}}$.
This term is 0 at $w=0^-$, goes negative at $w=-\psi$, and
will always be single peaked even for other bell-shaped
solitary potentials (for example, $\sech^2(x/\delta)$ and $\sech^4(x/\delta)$,
see Figure 4 of \cite{Turikov84} for the special case of empty-centered
EHs). Although the peak location may vary, it will not be at the end
points, 0 and $-\psi$.
The second term arising from the integral of the passing
electron distribution decreases monotonically from $w=0^-$ to $w=-\psi$.
The end point behavior of the two terms implies that
$f_{tr}(w=0^-) > f_{tr}(w=-\psi)$.
Combining the behavior of the two terms in $f_{tr}$, it can be concluded that
$f_{tr}(0>w\geq -\psi)\geq f_{tr}(w=-\psi)$.
This feature of $f_{tr}$ is essential in making
a solitary pulse, and it manifests itself at the peak of the potential as
two counterstreaming beams.
For the solution to be physical, $f_{tr}$ has to be nonnegative.
$f_{tr}(\psi, \delta, w=-\psi)\geq 0$ suffices this requirement,
yielding the width-amplitude ($\delta$-$\psi$) relation,
\begin{equation}
\delta \geq \left[\frac{2 \sqrt{\psi}(2\ln4-1)}{\sqrt{\pi}e^{\psi}
[1-\erf(\sqrt{\psi})]}\right]^{1/2}. \label{WA}
\end{equation}
The equal sign in Ineq. \ref{WA} corresponds to the case of empty-centered EHs
studied by Turikov \cite{Turikov84}.
We plot Ineq. \ref{WA} in Figure \ref{f2}.
A point in the shaded region represents an allowed EH with a given $\psi$ and
$\delta$. The shaded region includes all of the
allowed $\psi$ and $\delta$ for the range of values shown.
For a fixed $\delta$, all $\psi\leq\psi_0$ are
allowed, where $\psi_0$ is such that
$f_{tr}(\psi_0,\delta,w=-\psi_0)=0$; while for a fixed $\psi$, all
$\delta\geq\delta_0$ are allowed,
where $\delta_0$ is such that $f_{tr}(\psi,\delta_0,w=-\psi)=0$.
The corresponding physical meaning is that empty-centered EHs give the largest
amplitude ($\psi_0$) for a fixed width, and the smallest width ($\delta_0$)
for a fixed amplitude as comparing to EHs that are not empty-centered.
Note that by distributing a finite number of
electrons at rest at the bottom of the potential energy trough,
the plasma supports smaller amplitude or larger width
pulses than the amplitude and width of empty-centered EHs.
This inequality relation is dramatically different from that of KdV
solitons whose width-amplitude relation is a one-to-one mapping.

With $f_p$ and $f_{tr}$, we can now calculate the passing
and trapped electron densities separately and obtain
\[
n_p(x)=\exp{(\phi)}\left[1-\erf(\sqrt{\phi})\right],
\]
\[
n_{tr}(x)={\frac{-\phi \left[1+2\ln(\phi/\psi)\right]}{\delta^2}}+1-
\exp{(\phi)}\left[1-\erf(\sqrt{\phi})\right].
\]
Another way to obtain $n_{tr}$ comes directly from
solving Eq. \ref{phi2} as the first term on the RHS is exactly
$n_{tr}$.
Solving Eq. \ref{phi2} for $n_{tr}$ is simpler, but
without the knowledge of $f_{tr}$, one is not guaranteed whether the
particular set of ($\psi$, $\delta$) is physically allowed ($f_{tr}\geq0$).
Note that even for $\phi\ll 1$, $n_p\sim 1-2\sqrt{\phi/\pi}$ is different from
the leading terms of a Boltzmann distribution, $e^{\phi}$. The physical
meaning is that under self-consistent interaction of electrons and the
solitary potential, electrons in the vicinity of the potential are
not in local thermal equilibrium, in contrast to the starting point
of local thermal equilibrium in obtaining the
thermal screening length \cite{Debye23}.

\begin{figure}
\epsfxsize=8.5cm
\epsffile{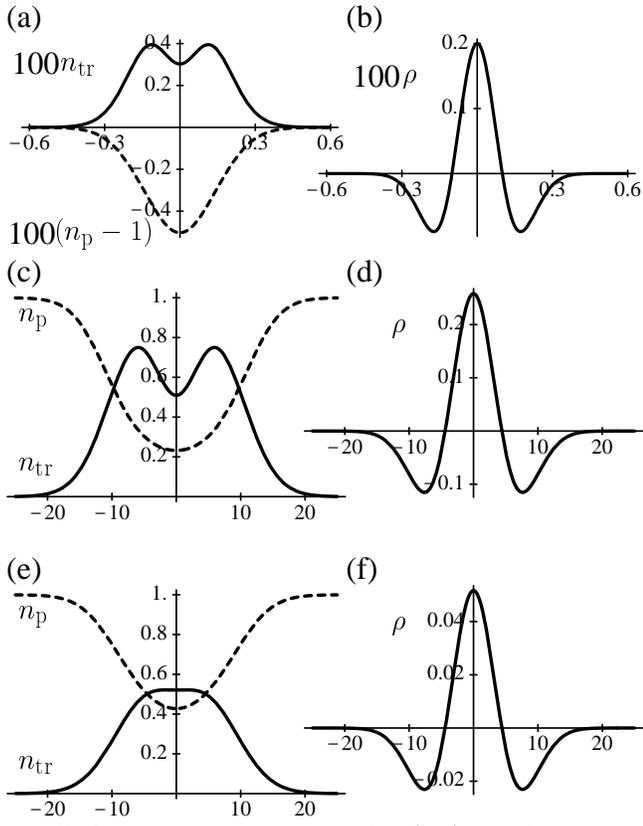}
\caption{Trapped electron density ($n_{tr}$), passing electron density ($n_p$),
and charge density ($\rho$) for ($\psi$,$\delta)=(2\times 10^{-5}, 0.1$) in
(a) and (b), ($\psi$,$\delta)=(5, 4.4)$ in (c) and (d), and ($\psi$,$\delta)=
(1, 4.4)$ in (e) and (f). }
\label{f3}
\end{figure}

To study the contributions from trapped and passing electrons to the
charge density ($-\partial^2 \phi/\partial x^2$) and how such contributions
are affected by various parameters, we show in Figure \ref{f3} plots
of $n_{tr}$ and $n_p$ and the charge density
$\rho$ as a function of $x$ for several values of $\psi$ and $\delta$.
Figure \ref{f3} (a) and (b) plot $100\times n_{tr}(x)$, $100\times
[n_p(x)-1]$, and $100\times \rho(x)$ for
($\psi$, $\delta$)=($2\times 10^{-5}$,$0.1$). For an ambient plasma with
$T_e=700 eV$ and $\lambda_D=100 m$ as found at ionospheric heights by FAST satellite
in the environment of BGK EHs \cite{Ergun99}, this case corresponds to $\psi=
1.4\times 10^{-2} V$ and $\delta=10 m$. As shown, in this weakly nonlinear case,
the maximum perturbation in $n_p$ is only $0.5\%$ and in $n_{tr}$ $0.4\%$.
The perturbation in the charge density $\rho$ is $\leq 0.2\%$, and occurs all
within one $\lambda_D$.

$n_p(x)$, $n_{tr}(x)$, and
$\rho(x)$ for ($\psi$, $\delta$)=(5, 4.4) in Figure \ref{f3} (c) and (d).
This choice corresponds to a point close to the curve
$f_{tr}(w=-\psi)=0$ in Figure 2 and is an extremely nonlinear case.
One can see that the total charge density perturbation goes
$\sim 10\%$ negative and $\sim 25\%$ positive, corresponding respectively
to electron density enhancement and depletion. 
With similar format, Figure \ref{f3} (e) and (f) plots a case with same $\delta$ and
$\psi=1$ to illustrate the change in $n_p$, $n_{tr}$, and $\rho$
of an EH with equal width but smaller amplitude. By locating this case
in Figure \ref{f2}, one notices that farther away from the $f_{tr}(w=-\psi)=0$
curve, the dip in $n_{tr}$ is filled up and the charge
density perturbation only increases to $5\%$ positive and $2\%$ negative.

These examples demonstrate how trapped electrons
produce negative charge density perturbations and passing electrons
positive charge density perturbations owing to the decrease in
their number density.
It is always true that $n_{tr}\geq 0$, since the number density cannot be
negative,
and therefore trapped electrons always contribute to negative charge density
regardless of the strength of the nonlinearity.
This result disagrees with
the picture that the positive core is due to
a deficit of deeply trapped electrons, and that this positive core is screened
out by the ambient electrons \cite{Schamel86}. It is also different from the
conclusion
that the trapped electrons are screened out by the resonant or nonresonant
passing electrons depending on the EH velocity in \cite{Krasovsky97}.

We now turn to the issue of minimum size of EHs.
From the illustrations of Figure \ref{f3}, one sees that
the charge density variation is the net balance
of the negative charge produced by trapped electrons and the positive charge
density produced by a depletion of passing electrons inside the solitary
potential.
The trapped electrons must distribute and oscillate in
such a way to yield the desired negative charge at the flanks
to shield out the positive core.
The entire solitary object is a self-consistent and self-sustained object with
zero total charge and does not require screening from the ambient plasma.
Thus, the size of EHs are not restricted to be greater than $\lambda_D$.

In summary, we demonstrated that the positive core of the EH is
shielded exclusively by trapped electrons oscillating between
their turning points and resulting in the excess negative charge.
A BGK EH consists of an electron density enhanced region and
depletion region, and this means that the total charge for a BGK EH is zero.
It thus follows that two separated EHs do not interact and the concept of
negative mass is not needed. There does not exist
a minimum size for BGK EHs since they do not rely on the thermal
screening from the ambient plasma. These features are independent of the particular
choice of potential profile and passing electron distribution, and also
independent of the strength of nonlinearity, because the only principle
upon which the arguments are built is conservation of charge
(the continuity equation).
The restriction to empty-centered EHs which was used previously is relaxed
to obtain a more general width-amplitude relation of
an inequality form.

Finally, note that the arguments and results we obtained for BGK
electron solitary waves can be
analogously applied to BGK ion solitary waves.                            

One of the authors (Chen) is grateful to Bill Peria for the discussions
on the electric field experiment onboard FAST satellite.
The research at the University of Washington
is supported in part by NASA grants NAG5-3170 and NAG5-26580.

\end{document}